\documentstyle [11pt]{article}
\newcommand{\f}{\frac}
\newcommand{\r}{\rho}
\newcommand{\p}{\partial}
\newcommand{\s}{\star}
\newcommand{\A}{{\cal A}}
\newcommand{\Dr}{{\cal D}}
\newcommand{\ap}{\approx}
\newcommand{\bc}{\begin{center}}
\newcommand{\ec}{\end{center}}
\newcommand{\be}{\begin{equation}}
\newcommand{\ee}{\end{equation}}

\newcommand{\M}{{\bf L}}
\newcommand{\k}{{\bf k}}
\newtheorem{prop}{Proposition}
\newtheorem{lemma}{Lemma}

\title{\huge{\bf Realization of $U_q(so(N))$ within the}\\
            {\bf Differential Algebra on ${\bf R}_q^N$}}

\author{{\bf Gaetano Fiore} \\ \\
{\it SISSA -- International School for Advanced Studies,} \\
{\it Strada Costiera 11, 34014 Trieste, Italy} \\ \\
{\it and}\\ \\
{\it Istituto Nazionale di Fisica Nucleare, Sezione di Trieste}}
\date{June 1993}
\begin{document}

\maketitle
\vspace{1.0cm}
\section*{\center Abstract}
We realize the Hopf algebra $U_{q^{-1}}(so(N))$ as an algebra of differential
operators on the quantum Euclidean space ${\bf R}_q^N$. The generators are
suitable q-deformed analogs of the angular momentum components on ordinary
${\bf R}^N$. The algebra $Fun({\bf R}_q^N)$ of functions on ${\bf R}_q^N$
splits into a direct sum of irreducible
vector representations of $U_{q^{-1}}(so(N))$; the latter are explicitly
constructed as highest weight representations.
\vskip4truecm

\newpage
\section{Introduction}
{}~~~~
One of the most appealing fact explaining
the present interest for quantum groups [1]
is perhaps the idea that they can be used to generalize the ordinary notion
of space(time) symmetry. This generalization is tightly coupled to a
radical modification of the ordinary notion of space(time) itself, and
can be performed through the introduction of a pair consisting of
a quantum group and the associated quantum space [2],[3].

The structure of a quantum group and of the corresponding quantum
space on which it coacts are intimely interrelated [2]. The
differential calculus
on the quantum space [4] is built so as to extend the covariant coaction
of the quantum group to derivatives. Here we consider the $N$-dimensional
quantum Euclidean space ${\bf R}_q^N$ and $SO_q(N)$ as the corresponding
quantum group; the Minkowski space and the Lorentz algebra could also be
considered, and we will deal with them elsewhere [5].

In absence of deformations,
a function of the space coordinates is mapped under
an infinitesimal $SO(N)$ transformation of the coordinates to a new one which
can be obtained through the action of some differential operators, the
angular momentum components. In other words the algebra $Fun({\bf R}^N)$ of
functions on ${\bf R}^N$ is the base space of a reducible representation
of $so(N)$, which we can call the regular (vector) representation of $so(N)$.
 It is interesting to ask whether an analog of this
fact occurs in the deformed case; in proper language, whether
$Fun({\bf R}_q^N)$ can
be considered as a left (or right) module of the universal
enveloping algebra $U_q(so(N))$, the latter being realized as some subalgebra
$U^N_q$ of the algebra of differential operators $Dif\!f({\bf R}^N_q)$ on
${\bf R}^N_q$.

In this paper we give a positive answer to this question. The result mimics
the classical (i.e. q=1) one: starting from the only $2N$ objects
$\{x^i,\p_j\}$
(the coordinates and derivatives, i.e. the generators of $Dif\!f({\bf R}_q^N)$)
with already fixed commutation and derivation relations, we end up with
a very economic way of realizing $U_{q^{-1}}(so(N))$ and its regular
vector representation (the fact that in this way we realize $U_{q^{-1}}(so(N))$
rather than $U_q(SO(N))$ is due to the choice that our differential
operators act from the left as usual, rather than from the right).
In this framework, the real structure
of $Dif\!f({\bf R}_q^N)$ induces the real structure of $U_{q^{-1}}(so(N))$.
This is
the subject of this work.

What is more, this approach makes inhomogeneous extensions
of $U_{q^{-1}}(so(N))$ and the study of the corresponding representation
spaces immediately at hand, without introducing any new generator:
it essentially suffices to add derivatives to the generators of
$U_{q^{-1}}(so(N))$ to find a realization of the q-deformed universal
enveloping algebra of the Euclidean algebra in $N$ dimensions [6], containing
$U_{q^{-1}}(so(N))$ as a subalgebra.
In fact this method was used in Ref. [7] to find the q-deformed Poincare'
Hopf algebra. In both cases the inhomogeneous Hopf algebra contains the
homogeneous one as a Hopf subalgebra, and we expect it to be the dual of
a inhomogeneous q-group constructed as a semidirect product in the sense
of Ref. [8].

The plan of the work is as follows. In section 2 we give preliminaries on the
quantum Euclidean space ${\bf R}_q^N$ and the differential algebra
$Dif\!f({\bf R}_q^N)$ on it. In section 3 we define a subalgebra
$U_q^N\subset Dif\!f({\bf R}_q^N)$ by requiring that its elements commute
with scalars, introduce two different sets
of generators for it and study the commutation relations of the second set.
In section 4 we find the commutation relations of these generators with
the coordinates and derivatives and derive the natural Hopf algebra structure
associated to $U_q^N$ (thought as algebra of differential operators on
$Fun({\bf R}_q^N)$); the Hopf algebra $U_q^N$ is then identified with
$U_{q^{-1}}(so(N))$. In Section 5 we find
that the q-deformed homogeneous symmetric spaces are the base spaces of
the irreducible representations of
$U_q^N$ in $Fun({\bf R}_q^N)$, and we show
that they can be explicitly constructed as highest weight representations.
When $q\in {\bf R}^+$ the representations are unitary and the hermitean
conjugation coincides with the complex conjugation in $Dif\!f({\bf R}_q^N)$.

We will treat by a unified notation odd and even $N$'s whenever it is possible,
and $n$ will be related to $N$ by the formulae $N=2n+1$ and $N=2n$
respectively.
We will assume that $q$ is not a root of unity.
Finally, we will often use the shorthand notation
$[A,B]_a:=AB-aBA$ ($\Rightarrow [\cdot,\cdot]_1=[\cdot,\cdot]$).

\section{Preliminaries}
{}~~~~

In this section we recollect some
basic definitions and relations characterizing the algebra $Fun({\bf R}_q^N)$
($O_q^N({\bf C})$ in the notation of [2]) of
functions on the quantum euclidean space
${\bf R}_q^N$, $N\ge 3$, (which is generated by the noncommuting
coordinates $x=(x^i)$), the ring $Dif\!f({\bf R}_q^N)$ of
differential operators on ${\bf R}_q^N$, the quantum group
$SO_q(N)$. In the first part we give a general overview of this matter;
in subsections 2.1, 2.2 we collect some more explicit formulae which we will
use in the following sections for explicit computations. In particular, in
subsection 2.2 we report on a very useful transformation [9] from the
$SO_q(N)$-covariant generators $x,\p$ of $Dif\!f({\bf R}_q^N)$ to completely
decoupled ones.
As in Ref. [9], index $i=-n,-n+1,...,-1,0,1,...n$ if
$N=2n+1$, and $i=-n,-n+1,...,-1,1,...n$ if $N=2n$.
For further details we refer the reader to [10], [9],[11].

{}~~~The braid matrix $\hat R_q:=||\hat R^{ij}_{hk}||$
for the quantum group $SO_q(N)$ is explicitly given by
\be
\hat R_q=q\sum \limits_{i\neq -i}e^i_i\otimes e^i_i +
\sum \limits_{i\neq j,-j\atop or~i=j=0}e^j_i\otimes e^i_j +
q^{-1}\sum \limits_{i\neq -i}e^{-i}_i\otimes e^i_{-i} +
\ee
\be
+(q-q^{-1})[\sum \limits_{i<j}e^i_i\otimes e^j_j -
\sum \limits_{i<j}q^{-\rho_i+\rho_j}e^{-j}_i\otimes e^j_{-i}],
\ee
where $(e^i_j)^h_k:=\delta^{ih}\delta_{jk}$. $\hat R_q$ is symmetric:
$\hat R^t=\hat R$.

{}~~~The q-deformed metric matrix $C:=||C_{ij}||$ is explicitly given by
\be
C_{ij}:=q^{-\rho_i}\delta_{i,-j},
\ee
where
\be
(\rho_i):=\cases {
(n-\f{1}{ 2},n-\f{3}{ 2},...,\f{1}{ 2},0,-\f{1}{ 2}...,\f{1}{ 2}-n)
{}~~~~~~~~~~if~N=2n+1 \cr
(n-1,n-2,...,0,0,...,1-n)~~~~~~~~~~~~~~~if~N=2n. \cr}
\ee
Notice that $N=2-2\r_n$ both for even and odd $N$.
$C$ is not symmetric and coincides with its inverse: $C^{-1}=C$.
Indices are raised
and lowered through the metric matrix $C$,  for instance
\be
a_i=C_{ij}a^j,~~~~~~a^i=C^{ij}a_j,
\ee

Both $C$ and $\hat R$
depend on $q$ and are real for
$q\in {\bf R}$. $\hat R$ admits the very useful decomposition
\be
\hat R_q = q {\cal P}_S - q^{-1} {\cal P}_A +q^{1-N}{\cal P}_1.
\ee
${\cal P}_S,{\cal P}_A,{\cal P}_1$ are the projection operators
onto the three eigenspaces of $\hat R$ (the latter have respectively dimensions
$\f{{N(N+1)}}{ 2}-1,\f{{N(N-1)}}{ 2},1$): they project the tensor
product $x\otimes x$ of the fundamental corepresentation $x$ of $SO_q(N)$
into the corresponding irreducible corepresentations (the symmetric modulo
trace, antisymmetric and trace, namely the q-deformed
versions of the corresponding ones of $SO(N)$). The projector ${\cal P}_1$ is
related to the metric matrix $C$ by ${\cal P}_{1~hk}^{~~ij}=\f{C^{ij}C_{hk}}{
Q_N}$; the factor $Q_N$ is defined by $Q_N:=C^{ij}C_{ij}$.
$\hat R^{\pm 1},C$ satisfy the relations
\be
[f(\hat R), P\cdot (C\otimes C)]=0~~~~~~~~~
f(\hat R_{12})\hat R^{\pm 1}_{23}\hat R^{\pm 1}_{12}=
\hat R^{\pm 1}_{23}\hat R^{\pm 1}_{12}f(\hat R_{23})
\ee
($P$ is the permutator: $P^{ij}_{hk}:=\delta^i_k\delta^j_h$ and $f$ is any
rational function);
in particular this holds for $f(\hat R)=\hat R^{\pm 1},{\cal P}_A,{\cal P}_S,
{\cal P}_1$.

{}~~~Let us recall that the unital
algebra $Dif\!f({\bf R}_q^N)$ of differential operators on the real quantum
euclidean plane ${\bf R}_q^N$ is defined as the space of formal series in the
(ordered)
powers of the $\{x^i\},\{\p_i\}$ variables, modulo the commutation relations
\be
{\cal P}_{A~hk}^{~~ij}x^h x^k =0.     ~~~~~~~~~~~~~
{\cal P}_{A~hk}^{~~ij}\p^h \p^k =0.
\ee
and the derivation relations
\be
\p_i x^j = \delta^j_i+q\hat R^{jh}_{ik} x^k\p_h.
\ee
The subalgebra $Fun({\bf R}_q^N)$ of `` functions '' on ${\bf R}_q^N$ is
generated by $\{x^i\}$ only. Below we will give the explicit form of these
relations.

For any function $f(x) \in Fun({\bf R}_q^N)$
$\p_i f$ can be expressed in the form
\be
\p_i f= f_i + f_i^j\p_j,~~~~~~~~~~f_i,f_i^j\in Fun({\bf R}_q^N)
\ee
(with $f_i,f^i_j$ uniquely determined)
upon using the derivation relations (9) to move step by step the derivatives
to the right of each $x^l$ variable of each term of
the power expansion of $f$, as far as the extreme right.
We denote $f_i$ by $\p_i f|$. This defines the action of $\p_i$ as a
differential operator
$\p_i:f\in Fun({\bf R}_q^N)\rightarrow \p_if|\in Fun({\bf R}_q^N)$: we will
say that $\p_if|$ is the `` evaluation '' of $\p_i$ on $f$. For instance:
\be
\p_i {\bf 1}|=0,~~~~~~~~\p^ix^j|=C^{ij},~~~~~~~~\p^ix^jx^k|=C^{ij}x^k+
q\hat R^{-1~ij}_{~~~hl}x^hC^{lk}
\ee
By its very definition, $\p_i$ satisfies the generalized Leibnitz rule:
\be
\p_i(fg)|=\p_i f|g+ {\cal O}_i^jf|\p_jg|,~~~~~~~~~f,g\in Fun({\bf R_q^N}),~~~~
{\cal O}^j\in Dif\!f({\bf R}_q^N)
\ee
(${\cal O}^jf|=f_i^j$). Any $D\in Dif\!f({\bf R}_q^N)$ can be considered as
a differential operator on $Fun({\bf R}_q^N)$ by defining its evaluation in a
similar way; a corresponding Leibnitz rule will be associated to it. In
section 4 we will consider as differential operators the angular momentum
components.

{}~~~If $q\in {\bf R}$ one can introduce an antilinear involutive
antihomomorphism
$*$:
\be
*^2=id~~~~~~~~~~~~~~~~~~~~~~~(AB)^*=B^*A^*
\ee
on $Dif\!f({\bf R}_q^N)$.  On the basic variables $x^i$  $*$ is
defined by
\be
(x^i)^*=x^jC_{ji}
\ee
whereas the complex conjugates of the derivatives $\p^i$ are not
combinations of the derivatives themselves. It is useful to introduce
barred derivatives $\bar \p^i$ through
\be
(\p^i)^*=-q^{-N}\bar \p^j C_{ji}.
\ee
They satisfy relation $(8)$ and the analog of (9) with
$q,\hat R$ replaced by $q^{-1},\hat R^{-1}$. These $\bar\p$ derivatives can
be expressed as functions of $x,\p$ [11], see formula (29).

{}~~~By definition a scalar $I(x,\p)\in Dif\!f({\bf R}_q^N)$ transforms
trivially
under the coaction associated to the quantum group of symmetry
$SO_q(N,{\bf R})$ [2]. Any scalar polynomial $I(x,\p)\in Dif\!f({\bf R}_q^N)$
of degree $2p$ in $x,\p$ is a combination of terms of the form
\be
I=(\eta_{\varepsilon_1})^{i_1}(\eta_{\varepsilon_2})^{i_2}...(\eta_
{\varepsilon_p})^{i_p}(\eta_{\varepsilon'_p})_{i_p}...(\eta_{\varepsilon'_2})
_{i_2}(\eta_{\varepsilon'_1})_{i_1},
\ee
where $\varepsilon_i,\varepsilon'_j=+,-$, $\eta_+:=x$ and $\eta_-:=\p$.
{}From here we see that no polynomial of odd degree in $\eta^i_
\varepsilon$ can be a scalar. One can show that any scalar polynomial
$I(x,\p)$
can be expressed as an ordered polynomial in two particular scalar
variables (see for instance Appendix C of [12]),
namely the square lenght  $x\cdot x$ and the laplacian
$\p\cdot \p$, which are defined in formulae (20) below.

We will use two types of q-deformed integers:
\be
[n]_q := \f{{q^n-q^{-n}}}{ {q-q^{-1}}}~~~~~~~~~(n)_q:=\f{q^n-1}{q-1};
\ee
both $[n]_q$ and $(n)_q$ go to $n$ when $n \rightarrow 1$.

\subsection{Some explicit formulae in terms of $x,\p$ generators}

For any `` vectors '' $a:=(a^i),b:=(b^i)$ let us define
\be
(a\s b)_j:=\sum\limits_{l=1}^j a^{-l}b_{-l}+
\cases{\f{q}{q+1}a^0b_0~~~~if~N~odd\cr 0~~~~~~~~~~~~~
if~N~even},~~~~~~~~~~0\le j\le n
\ee
\be
\A^j(a,b):=a^jb^{-j}-a^{-j}b^j-(q^2-1)q^{-\r_j-2}(a\s b)_{j-1},~~~~~~~~~j\ge 1
\ee
\be
(1+q^{-2\r_j})(a\cdot b)_j:=\sum\limits_{l=-j}^j a^lb_l~~~~~~~
{}~~~~~~~~~~~~0< j\le n
\ee
(when this causes no confusion we will also use the notation
$a\cdot b:=(a\cdot b)_n$).
Then it is easy to verify that
\be
(a\cdot b)_j=(a\s b)_j+\f{\sum\limits_{l=1}^j\A^l(a,b)q^{-\r_l}}{1+q^{-2r_j}}
\ee
Note that the preceding four formulae make sense for any $n\ge j$ and do not
formally depend on $n$.

Relations (8), (9) defining $Dif\!f({\bf R}_q^N)$ amount respectively to
\be
x^ix^j=qx^jx^i,~~~~~~~~~~\p^i\p^j=q\p^j\p^i~~~~~~~~~~i<j,
\ee
\be
\A^i(x,x)=0,~~~~~~~\A^i(\p,\p)=0           ~~~~~~~~~~i=1,2,...,n
\ee
and
\be
\cases{
\p_kx^j=qx^j\p_k-(q^2-1)q^{-\r_j-\r_k}x^{-k}\p_{-j},~~~
{}~~~~~~~~~~~~~~~~~~j<-k,j\neq k \cr
\p_kx^j=qx^j\p_k
{}~~~~~~~~~~~~~~~~~~~~~~~~~~~~~~~~~~~~~~~~~~~~~~~~~~~~~~~j>-k,j\neq k \cr
\p_{-k}x^{k}=x^k\p_{-k},~~~~~~~
{}~~~~~~~~~~~~~~~~~~~~~~~~~~~~~~~~~~~~~~~~~~~~~~~~~~~~~~~~~k\neq 0  \cr
\p_ix^i=1+q^2x^i\p_i+(q^2-1)\sum\limits_{j>i}x^j\p_j,
{}~~~~~~~~~~~~~~~~~~~~~~~~~~~~~~~~~~~i>0 \cr
\p_ix^i=1+q^2x^i\p_i+(q^2-1)\sum\limits_{j>i}x^j\p_j
-q^{-2\r_i}(q^2-1)x^{-i}\p_{-i},
{}~~~~~~~i<0 \cr
\p_0x^0=1+qx^0\p_0+(q^2-1)\sum\limits_{j>0}x^j\p_j,
{}~~~~~~~~~~~~~~~~~~~~~~~(only~for~N~odd). \cr}
\ee
Here are some useful formulae (sum over $l$ is understood):
\be
\p^i (x\cdot x)_n=q^{2\r_n}x^i+q^2(x\cdot x)_n\p^i~~~~~~~
(\p\cdot \p)_n x^i=q^{2\r_n}\p^i+q^2x^i(\p\cdot\p)_n
\ee
\be
(x^l\p_l) x^i=x^i+q^2x^i x^l\p_l +(1-q^2)(x\cdot x)_n \p^i~~~~~~~
\p^i (x^l\p_l)=\p^i+q^2(x^l\p_l) \p^i+(1-q^2)x^i(\p\cdot \p)_n.
\ee

In Ref. [9],[11] the dilatation operator $\Lambda_n$
\be
\Lambda_n(x,\p):=1 +(q^2-1)x^i\p_i+q^{N-2}(q^2-1)^2(x\cdot x)(\p\cdot\p)
\ee
was introduced; it fulfils the relations
\be
\Lambda_n x^i=q^2x^i\Lambda_n,~~~~~~~~~~\Lambda_n\p^i=q^{-2}\p^i\Lambda_n,
\ee
Then one can prove [11] that
\be
\bar \p^k=\Lambda_n^{-1}[\p^k+q^{N-2}(q^2-1)x^k(\p\cdot\p)]
\ee
In the sequel we will also need the operator
\be
{\cal B}_n:=1+q^{N-2}(q^2-1)(x\cdot\p)
\ee
it is easy to show that it is the only operator of degree one in $x^i\p^j$
satisfying the relations
\be
{\cal B}_n(x\cdot x)=q^2(x\cdot x)B,~~~~~~~~~~~~
{\cal B}_n(\p\cdot\p)=q^{-2}(\p\cdot\p) B_n;
\ee
Under complex conjugation
\be
{\cal B}_n^*=q^{-N}{\cal B}\Lambda_n^{-1}
{}~~~~~~~~~~~~\Lambda_n^*=q^{-2N}\Lambda^{-1}_n~~~~~~if~~q\in {\bf R}^+.
\ee
In the sequel we will drop the index $n$ in ${\cal B}_n,\Lambda_n$ when this
causes no confusion.

\subsection{Decoupled generators of $Dif\!f({\bf R}_q^N$)}

{}~~~In Ref. [9] it is shown that there exists a natural embedding of
$Dif\!f({\bf R}^{N-2}_q)$ into $Dif\!f({\bf R}^N_q)$. In next sections we will
see
that it naturally induces an embedding of $U_q(so(N-2))$ into
$U_q(so(N))$.
We just need to do the change of generators of $Dif\!f({\bf R}^N_q)$
$(x^i,\p^j)\rightarrow (X^i,D^j)$ ($|i|,|j|\le n$), with
\be
\cases{
x^i=\mu_n^{\f{1}{2}}X^i,~~~~~~~~\p_i=\mu_n^{\f{1}{2}}D_i,~~~~~~~~|i|<n \cr
x^n=X^n~~~~~~~~~~\p_n=D_n \cr
x^{-n}=\Lambda_n^{\f{1}{2}}\mu_n^{\f{-1}{2}}X^{-n}-q^{-2-\r_n}(q^2-1)(X\cdot
X)_{n-1}D_n \cr
\p_{-n}=q^{-1}\Lambda_n^{\f{1}{2}}\mu_n^{\f{-1}{2}}D_{-n}-q^{-2-\r_n}
(q^2-1)X^n(D\cdot D)_{n-1} \cr}
\ee
and
\be
\mu_n:=\mu(X^n,D_n):=D_nX^n-X^nD_n=1+(q^2-1)X^nD_n.
\ee
Then the variables $X^i,D^j$, ($|i|,|j|\le n-1$) satisfy the commutation and
derivation relations (8),(9) for $Dif\!f({\bf R}_q^{N-2})$, whereas
$$
[X^{\pm n},X^i]=0~~~~~~~~[X^{\pm n},D_i]=0~~~~~~~~
[D_{\pm n},X^i]=0~~~~~~~~[D_{\pm n},D_i]=0,
$$
\be
[D_{\pm n},X^{\mp n}]=0~~~~~~~[D_n,D_{-n}]=0~~~~~~~[X^n,X^{-n}]=0
\ee
and
\be
D_nX^n=1+q^2X^nD_n~~~~~~~~~~~~~~D_{-n}X^{-n}=1+q^{-2}X^{-n}D_{-n}.
\ee
As a direct consequence of the previous relations, $\mu_n$ commutes with
all the $X,D$ variables, except $X^n,D_n$ themselves:
\be
\mu_nX^n=q^2X^n\mu_n,~~~~~~~~~\mu_nD_n=q^{-2}D_n\mu_n.
\ee
The dilatation operator
$\Lambda_n$ in terms of $X^i,D^j$ variables reads
$\Lambda_n(x,\p)=\Lambda_{n-1}(X,D)\mu_n\mu_{-n}$, where
$\mu_{-n}:=(D_{-n}X^{-n}-X^{-n}D_{-n})^{-1}$ and $\Lambda_{n-1}(X,D)$ depends
only on $X^i,D_j$ ($|i|,|j|\le n-1$) as dictated by formula
$(27)$ (after the replacement $n\rightarrow n-1$).

For odd $N$ it is convenient to start the chain of embeddings from the
`` differential algebra of the quantum line '' $Dif\!f({\bf R}_q^1)$
generated by $x^0,\p_0$ satisfying the relation
\be
\p_0x^0=1+qx^0\p_0.
\ee
For $N$ even, it is convenient to start the chain from the differential algebra
$Dif\!f({\bf R}_q^2)$ of two commuting quantum lines; it is generated by the
four variables $x^{\pm 1},\p_{\pm 1}$ all commuting  with each-other,
except for the relations
\be
\p_{\pm 1}x^{\pm 1}=1+q^2x^{\pm 1}\p_{\pm 1}.
\ee
Summing up, we have the two chains of embeddings
\be
Dif\!f({\bf R}_q^h)\hookrightarrow Dif\!f({\bf R}_q^{h+2})\hookrightarrow
Dif\!f({\bf R}_q^{h+4})\hookrightarrow.........
\ee
where $h=\cases{1~~~for~odd~N's\cr 2~~~for~even~N's\cr}$.

{}From the abovementioned embeddings it trivially follows the important

\begin{prop}
\be
F(x^i,\p_j)=0~~~in~Dif\!f({\bf R}_q^{N-2})~~~\Rightarrow~~~
F(X^i(x^h,\p_k),D_j(x^h,\p_k))=0~~~in~Dif\!f({\bf R}_q^N),
\ee
with $|i|,|j|\le n-1,~~~~|h|,|k|\le n$.
In the LHS $x^i,\p_j$ are the $(x,\p)$-type generators for
$Dif\!f({\bf R}_q^{N-2})$,
in the RHS $X^i,D_j$ and $x^h,\p_k$ are respectively $X,D$- and
$(x,\p)$-type generators for $Dif\!f({\bf R}_q^N)$, and $F$ for our purposes
will be some polynomial
function in the variables
$x,\p,\mu_{n-1}^{\pm \f 12} \Lambda_{n-1}^{\pm \f 12}$.
\end{prop}

Let us introduce variables $\chi^i,{\cal D}_i$, $i\in {\bf Z}$, such that
\be
{\cal D}_i\chi^i=1+ a\chi^i{\cal D}_i~~~~~~~~~~a=\cases{q^2~~~~~if~i>0~~~or
{}~N~even~and~i=-1\cr q~~~~~~~~~~if~i=0\cr q^{-2}~~~~~~~~otherwise\cr},
\ee
\be
[\eta,\xi]=0~~~~~~~if~~~~\eta=\chi^i,{\cal D}_i,~~~\xi=\chi^j,{\cal D}_j
{}~~~~with~~i\neq j.
\ee
By iterating the transformation (33) one arrives precisely at generators of
$Dif\!f({\bf R}_q^N)$ of the type $\chi^i,{\cal D}_i$ with $|i|\le n$
(and $i\neq 0$ when $N$ is even), by identifying
\be
X^{\pm n}=\chi^{\pm n},~~~~D_{\pm n}={\cal D}_{\pm n},~~~~~~~~
X^{n-1}=\chi^{n-1},~~~~D_{n-1}={\cal D}_{n-1}~~~....
\ee
We generalize the  definition (34) in the following way
\be
\cases{(\mu_{\pm i})^{\pm 1}:=\Dr_{\pm i}\chi^{\pm i}-\chi^{\pm i}\Dr_{\pm i}=
1+(q^{\pm 2}-1)\chi^{\pm i}\Dr_{\pm i}~~~~~~~~~i>0,~~~except~when~N~even~and~
i=1;\cr
\mu_{\pm1}:=\Dr_{\pm 1}\chi^{\pm 1}-\chi^{\pm 1}\Dr_{\pm 1}=
1+(q^2-1)\chi^{\pm 1}\Dr_{\pm 1}~~~~~~~~~~~~~~~~~~~when~N~even; \cr
(\mu_0)^{\f 12}
:=\Dr_0\chi^0-\chi^0\Dr_0= 1+(q-1)\chi^0\Dr_0={\cal B}_0
{}~~~~~~~~~~~~~~~~~~~when~N~odd. \cr}
\ee
Consequently,
\be
[\mu_i,\mu_j]=0~~~~~~\mu_i\chi^j=\chi^j\mu_i\cdot\cases{q^2~~~if~~i=j\cr
1~~~if~~i\neq j\cr},~~~~~~~~~~~~\mu_i\Dr_j=\Dr_j\mu_i\cdot\cases{q^{-2}
{}~~~if~~i=j\cr 1~~~if~~i\neq j\cr}.
\ee
and $\Lambda_n=\prod\limits_{i=-n}^n \mu_i$.
In terms of $X,D$ and $\chi,{\cal D}$ variables the square lenght $x\cdot x$
and the laplacian $\p\cdot\p$ take respectively the forms
\be
x\cdot x=\Lambda_n^{\f 12}\mu_n^{-\f 12}X^nX^{-n}q^{\r_n}+q^{-2}(X\cdot
X)_{n-1}
=\sum\limits_{i=1}^n\Lambda_i^{\f 12}\mu_i^{-\f 12}\chi^i\chi^{-i}q^{\r_i-
2(n-i)}+\cases{0~~~~~~~~~~~~~if~~N=2n \cr \f{q^{-2n+1}}{q+1}\chi^0\chi^0~~~~
if~~N=2n+1\cr}
\ee
$$
\p\cdot \p=\Lambda_n^{\f 12}\mu_n^{-\f 12}D^nD^{-n}q^{\r_n-1}+q^{-2}
(D\cdot D)_{n-1}=\sum\limits_{i=1}^n\Lambda_i^{\f 12}\mu_i^{-\f 12}
{\cal D}_i{\cal D}_{-i}q^{\r_i-1-2(n-i)}
$$
\be
+\cases{0~~~~~~~~~~~~~~~~if~~N=2n \cr
\f{q^{-2n+1}}{q+1}{\cal D}_0{\cal D}_0~~~~~~~if~~N=2n+1\cr}
\ee

\section{The U.E.A. of the angular momentum on ${\bf R}_q^N$}
{}~~~~

Inspired by the classical (i.e. q=1) case, we give the following

{\bf Definition:} the universal enveloping
algebra $U_q^N$ of the angular momentum on ${\bf R}_q^N$ is the subalgebra
of $Dif\!f({\bf R}_q^N)$ whose elements commute with any
scalar $I(x,\p)\in Dif\!f({\bf R}_q^N)$. Since any such $I$ can be expressed as
a
function of the laplacian and of the square lenght $x\cdot x,\p\cdot \p$,
our definition amounts to
\be
U_q^N:=\{u\in Dif\!f({\bf R}_q^N):~~~~[u,x\cdot x]=0=[u,\p\cdot\p]\}
\ee
In the next two subsections we consider two sets of generators of $U_q^N$
(actually we will prove in Appendix B that any $u\in U_q^N$ can be
expressed as a function of them). The generators of the first set
transform in the same way as the products $x^ix^j$ under the coaction,
since (up to a scalar) they are q-antisymmetrized products
of $x,\p$ variables, but have rather complicated commutation
relations; nevertheless Casimirs have a very compact expressions in terms of
them. The generators of the second set have a quite simple form in terms
of $\chi,{\cal D}$ variables and are much more useful for practical purposes,
since they have simple commutation relations and are directly connected with
the Cartan-Weyl generators of $U_{q^{-1}}(so(N))$

\subsection{The set of generators $\{L^{ij},B\}$}
{}~~~~

Keeping the
classical case in mind, where the angular momentum components are
antisymmetrized products $x^i\p^j-x^j\p^i$ of coordinates and
derivatives, we try with the q-deformed antisymmetrized products
\be
{\cal L}^{ij}:={\cal P}_{A~hk}^{~~~ij}x^h\p^k=-q^{-2}
{\cal P}_{A~hk}^{~~~ij}\p^hx^k.
\ee
{}From relations (25),(8) it follows that
\be
{\cal L}^{ij}x\cdot x=q^2x\cdot x{\cal L}^{ij}~~~~~~~~
{\cal L}^{ij}\p\cdot\p=q^{-2}\p\cdot\p{\cal L}^{ij}.
\ee
This implies that ${\cal L}^{ij}$ commutes only with scalars having
natural dimension $d=0$. This shortcoming can be cured by introducing a scalar
$S\in Dif\!f({\bf R}_q^N)$ with natural dimension $d=0$ and such that
$Sx\cdot x=q^{-2}x\cdot xS$, $S\p\cdot\p=
q^2\p\cdot\p S$; then by defining $L^{ij}:={\cal L}^{ij}S$ we get
\be
[L^{ij},I]=0;
\ee
$L^{ij}$ are therefore candidates to the role of angular momentum components.
The simplest choice is to take $S=\Lambda^{-\f 12}$, as we did in Ref. [12],
and will be adopted in the sequel.

Starting from commutation relations for the ${\cal L}^{ij}$'s we get
corresponding relations for the $L^{ij}$'s by multiplying them by a suitable
power of $\Lambda^{-\f 12}$. In fact, it is clear that the former must be
homogeneous
in ${\cal L}$'s to be consistent with (51). Nevertheless, commutation
relations including factors such as ${\cal L}^{ij}{\cal L}^{-j,l}$
cannot be of this form. In fact, performing the derivations $\p^{-j} x^j$
according to rules (9) one
lowers by 1 the degree in $x\p$ of some terms; this can be taken into account
only by considering homogeneous relations both in ${\cal L}^{ij}$'s and
${\cal B}$ (${\cal B}$ was defined in (30)), since ${\cal B}$ is the only other
$1^{st}$ degree polynomial in $x^i\p_j$ with  the same scaling law (51) as
${\cal L}^{hk}$.
Summing up, we expect homogeneous commutation relations  in the $L^{ij}$'s and
$B:={\cal B}\Lambda^{-\f 12}$. $B$ is not really an independent generator,
as we will see below. Therefore, the alternative choice $S:={\cal B}^{-1}$
(as considered in ref. [13]) would yield the same algebra.

{\bf Remark 1:} When $q=1$ ${\cal B}=1=\Lambda$ and
$L^{ij}$ reduce to the classical
`` angular momentum '' components, i.e. to generators of $U(so(N))$ (note
that they are expressed as functions of the non-real coordinates $x^i$ of
${\bf R}^N$ and of the corresponding derivatives). In this limit one can
take as generators of the Cartan
subalgebra the $L^{i,-i}$'s, as ladder operators corresponding to
positive (resp. negative) roots the $L^{jk}$'s with $|j|<|k|$ and $k>0$
(resp. $k<0$), as ladder operators corresponding to simple roots
the $L^{1-i,i}$'s together with $L^{j,2}$ ($i=2,...,n$, and
$j=\cases{0~~~if~N=2n+1\cr 1~~~if~N=2n\cr}$).
A Chevalley basis is formed by the set of triples
$\{(L^{1-i,i},L^{-i,i-1},L^{i,-i}-L^{i-1,1-i}),~~~i=1,...,n\}$
if $N=2n+1$ (here $ L^{0,0}=0$) and
$\{(L^{1,2},L^{-2,-1},L^{2,-2}+L^{1,-1}),
(L^{1-i,i},L^{-i,i-1},L^{i,-i}-L^{i-1,1-i}),~~~i=2,...,n\}$ if $N=2n$.
The correspondence with spots in the Dynkin diagrams of the classical series
${\bf B}_n,{\bf D}_n$ is shown in figure 1.

{}~

{\bf Remark 2:} One could work with $\bar \p$ instead of $\p$ derivatives
and define $\bar L^{ij}:={\cal P}^{~~ij}_{A~hk}x^h\bar\p^k\Lambda^{\f 12}$
$\bar B_n:=\bar {\cal B}_n\Lambda^{\f 12}$,
where $\bar {\cal B}_n:=1+(q^{-2}-1)(x\cdot\bar \p)$. But using formulae
(15),(28),(32) one shows
that $q^{-1}\bar L^{ij}=q L^{ij}$. In the language of Ref. [12] this means
that the angular momentum in the barred and unbarred representation
essentially coincide.

{}~

Instead of the $N$ linearly dependent operators ${\cal L}^{-i,i}$ one can
use their $n$ linearly independent combinations
\be
{\cal L}^i:=\A^i(x,\p),  ~~~~~~~~~i=1,2,...,n.
\ee
As for the operators ${\cal L}^{ij},~i\neq -j$, for simplicity we will
renormalize them as follows
\be
{\cal L}^{ij}:=(1+q^2){\cal P}_{A~hk}^{~~ij}x^i\p^j=(x^i\p^j-qx^j\p^i),
{}~~~~~~i<j,~~~~~~~~~~
{\cal L}^{ij}=-q {\cal L}^{ji},~~~~~~~i>j.
\ee

The scalar $(L\cdot L)_n:=L^{ij}L_{ji}$ commutes with any $L^{ij}$ and
reduces (up to a factor) to the classical square angular momentum when q=1.
We will call this casimir the (q-deformed) square angular momentum.
Higher order Casimirs can be obtained by forming nontrivial independent
scalars out of $j-th$ powers ($j>2$) of the $L$'s,
\be
(\underbrace{L\cdot L\cdot ...L}_{j~times})_n
:=L^{i_1i_2}L_{i_2}^{~i_3}L_{i_3}^{~i_4}...L_{i_j,i_1},
\ee
for the same values of j as in the classical case.

\begin{prop}
The following important relation connects $\Lambda, {\cal B}$ and
${\cal L}\cdot {\cal L}$:
\be
\Lambda_n=({\cal B}_n)^2-\f{(q^2-1)(q^2-q^{-2})}{(1+q^{2\r_n})(1+q^{-2\r_n-2})}
({\cal L}\cdot {\cal L})_n.
\ee
\end{prop}
$Proof$. Using formulae (7),(6),(8) one can easily show [12] that
\be
({\cal L}\cdot {\cal L})_n=\alpha_N(q)x^i\p_i+\beta_N(q)x^ix^j\p_j\p_i+
\gamma_N(q)(x^ix_i)(\p^i\p_i),
\ee
where
\be
\alpha_N(q):=\f{(q^{2-\f{N}{ 2}}+q^{\f{N}{ 2}-2})(q^{1-N}-q^{N-1})}{
(q^{1-\f{N}{ 2}}+q^{\f{N}{ 2}-1})(q^{-2}-q^2)}
\ee
\be
\beta_N:=\f{q^3+q^{N-1}}{(1+q^{2-N})(q+q^{-1})}~~~~~~~~~~~
\gamma_N:=-\f{(q^{5-N}+q)(1+q^{-N})}{(1+q^{2-N})^2(q+q^{-1})}.
\ee
Performing derivations in ${\cal B}^2$ according to formula (26) we realize
that
the RHS of formula (56)
gives $\Lambda$ as defined in formula (27). $\diamondsuit$

As a consequence, $B^2$ is not an independent generator, as anticipated,
but depends on $L\cdot L$.

When $q\in {\bf R}$ from formulae (14),(15),(28),(32) it follows that under
complex conjugation
\be
(L^{i,j})^*=q^{\r_i+\r_j}L^{-j,-i};
\ee
this implies in particular that $L\cdot L$, the other casimirs
and the $L^i$'s  are real.
Moreover, it is easy to show that all the $L^i$'s commute with each other,
as the ${\cal L}^i$'s do.

The basic commutation relations between $L^{ij},B$ are quadratic in these
variables but rather complicated and we won't give them here.

\subsection{The set of generators $\{\M^{ij},k^i\}_{i\neq j}$}
{}~~~~

On the contrary,
the new generators defined below admit very simple commutation relations,
allowing a straightforward proof of the isomorphism
$U_q^N\ap U_{q^{-1}}(so(N))$. It is convenient to use $\chi,{\cal D}$ variables
to define and study them. The definitions of $\M^{ij},\k^i$
involve only $\chi^l,{\cal D}_m$ variables with $|l|,|m|\le J:=max\{|i|,|j|\}$,
so that in terms of these variables it makes sense for any $n\ge J$. Hence,
it trivially follows the embedding $U_q^N \hookrightarrow U_q^{N+2}$,
 since,
as we will show in Appendix B, $\M^{ij},\k^i$ generate $U_q^N$.

\begin{prop}
The elements
\be
\k^i:=\mu_i\mu_{-i}^{-1}\in Dif\!f({\bf R}_q^N)
{}~~~~~~~0<i\le n
\ee
belong to $U_q^N$ and commute with each other.
\end{prop}
$Proof$. The thesis is a trivial consequence of formulae (46),(47),(48).
$\diamondsuit$

We will call the subalgebra
generated by $\k^i$ the `` Cartan subalgebra '' $H_q^N\subset U_q^N$.

In Appendix A we show that the elements $\k_i\in U_q^N$ can be expressed as
functions of  $B,L^{ij}$.

Now we define the generators $\M^{ij}\in U_q^N$,
which correspond to roots. Since the generators of $U_q^{N-2}$ belong also
to $U_q^N$ (in the sense of the abovementioned embedding), we can stick to
the definition of the new
generators, i.e. the ones belonging to ($U_q^N-U_q^{N-2}$). For this purpose
it is convenient to use the $X,D$ variables of $Dif\!f({\bf R}_q^N)$.

{\bf Definition}:
\be
\cases{\M^{ln}:=q^{-2}\Lambda^{-\f 12}_n\mu_{-n}[D^l,(X\cdot X)_{n-1}]D^n-
\mu_n^{-\f 12}X^nD^l,~~~~~~~~~~~~~~~~~~(positive~ roots)\cr
\M^{-nl}:=q^{-1}\Lambda^{-\f 12}_n\mu_{-n}X^{-n}[(D\cdot D)_{n-1},X^l]-
\mu_n^{-\f 12}X^lD^{-n},~~~~~~~~~~~(negative~ roots)\cr}~~~~~~~~~~~|l|<n.
\ee
In particular, it is easy to show that the complete list of generators
corresponding to simple roots of $U_q^N$ (i.e. the ones with
indices as prescribed in Remark 1) in terms of $\chi,{\cal D}$ variables reads
\be
\cases{\M^{1-k,k}:=\mu_k^{-\f 12}\left[q^{2\r_k}(\mu_{-k}\mu_{k-1})^{\f
12}\chi^
{1-k}\Dr^k - \chi^k\Dr^{1-k}\right]~~~~~~~~~~j\le k\le n,~~~~~~
j=\cases{2~~~if~~N=2n+1\cr 3~~~if~~N=2n\cr}\cr
\M^{01}:=(\mu_1)^{-\f 12}\left[q^{-2}(\mu_{-1})^{\f 12}\chi^0\Dr^1-
\chi^1\Dr^0\right]~~~~~~~~~~~~~~~~~~~~~~~~~~~~if~~N=2n+1, \cr
\M^{\pm 1,2}:=\mu_2^{-\f 12}\left[q^{-2}(\mu_{-2}\mu_{\mp 1})^{\f 12}
(\mu_{\pm 1})^{-\f 12}
\chi^{\pm 1}\Dr^2 - \chi^2\Dr^{\pm 1}\right]~~~~~~~~~~~~if~~N=2n;\cr}
\ee
the list of corresponding negative Chevalley partners is given by
\be
\cases{\M^{-k,k-1}:=\mu_k^{-\f 12}\left[q^{2\r_k-1}(\mu_{-k}\mu_{k-1})
^{\f 12}\chi^{-k}\Dr^{k-1} - \chi^{k-1}\Dr^{-k}\right]~~~~~~~~~~2\le k\le n,\cr
\M^{-1,0}=(\mu_1)^{-\f 12}\left[\mu_{-1}^{\f 12}\chi^{-1}\Dr^0-\chi^0\Dr^{-1}
\right]~~~~~~~~~~~~~~~~~~~~~~~~~~~~~~~if~~N=2n+1,\cr
\M^{-2,\pm 1}:=\mu_2^{-\f 12}\left[q^{-1}(\mu_{-2}\mu_{\pm 1})^{\f 12}
(\mu_{\mp 1})^{-\f 12}
\chi^{-2}\Dr^{\pm 1} - \chi^{\pm 1}\Dr^{-2}\right]~~~~~~~~~~~~~~~if~~N=2n.\cr}
\ee
Note that when $N=2n+1$ $L^{0\pm 1}=(\k^1)^{\f 12}\M^{0\pm 1}$, when $N=2n$
$L^{\pm 1,2}=(\k^2)^{\f 12}(\k^1)^{\pm\f 12}\M^{\pm 1,2}$,
$L^{-2,\pm 1}=q(\k^2)^{\f 12}(\k^1)^{\mp\f 12}\M^{-2,\pm 1}$.
In appendix A we show that the simple roots and their Chevalley partners
are functions of $L^{ij},B$.

\begin{prop}
: $\M^{ln},\M^{-n,l}\in U_q^N.$
\end{prop}
$Proof$. In terms of $X,D$ variables, formulae (47),(46),(42) yield
$$
[\M^{ln},(x\cdot x)_n]=\left[q^{-2}\Lambda_n^{-\f 12}\mu_{-n}[D^l,(X\cdot X)_
{n-1}]D^n,\Lambda_n^{\f 12}\mu_{-n}^{-\f 12}X^nX^{-n}q^{\r_n}\right]-
[\mu_n^{-\f 12}X^nD^l,q^{-2}(X\cdot X)_{n-1}]
$$
\be
=q^{\r_n-2}\mu_{-n}\mu_n^{-\f 12}[D^l,(X\cdot X)_{n-1}][D^n,X^{-n}]X^n-
q^{-2}\mu_n^{-\f 12}X^n[D^l,(X\cdot X)_{n-1}]=0,
\ee
and formulae (48),(46),(42) yield
$$
[\M^{ln},(\p\cdot \p)_n]=-q^{\r_n}\mu_n^{-1}\Lambda_n^{\f 12}[X^n,D^{-n}]_
{q^{-2}}D^lD^n+q^{-4}\Lambda_n^{-\f 12}\mu_{-n}D^n\left[[D^l,(X\cdot X)_{n-1}],
(D\cdot D)_{n-1}\right]_{q^{-2}}
$$
\be
=q^{2\r_n-2}\mu_n^{-1}\Lambda_n^{\f 12}D^lD^n-q^{-6}\Lambda_n^{-\f 12}
\mu_{-n}D^n\left[D^l,\Lambda_{n-1}\f{q^{4+2\r_n}}{q^2-1}\right]=0
\ee
(here we have used the identity
$[\p\cdot\p,x\cdot x]_{q^2}=\f{q^{2+2\r_n}}{q^2-1}(\Lambda_n-1)$);
namely $\M^{ln}\in U_q^N$. Similarly one proves that $\M^{-nl}\in U_q^N$.
$\diamondsuit$.

\begin{lemma}
\be
[\M^{hn},\p_n]_q=\p^h,~~~~~~~~~~~[\M^{-n,h},x^n]_{q^{-1}}=-q^{\r_n}x^h
{}~~~~~~~~~~~~~~~~~~~~~~~~~|h|<n,
\ee
\be
[\p^{n-1},\M^{1-n,n}]_{q^{-1}}=q^{\r_n-1}\p^n,~~~~~~~~[\M^{-n,n-1},x^{1-n}]
_{q^{-1}}=q^{\r_l}x^{-n}~~~~~~~~~~~~~~~~~~~~~~~~~n>1,
\ee
\be[\p^0,\M^{01}]=q^{-1}\p^1~~~~~~~~~~~[\M^{-10},x^0]=x^{-1}
{}~~~~~~~~~~~~~~~~~~~~~~~~~if~~N=3.
\ee
\end{lemma}
$Proof$. For the proof see Proposition 11  of next section and the remark
following it. $\diamondsuit$

The following proposition allows to construct all the roots starting from
the Chevalley ones.
\begin{prop}
The following relations hold in $Dif\!f({\bf R}_q^N)$:
\be
[\M^{-jl},\M^{-lk}]_q=q^{\r_l}\M^{-j,k}~~~~~~~~~~[\M^{-kl},\M^{-l,j}]
_q=q^{\r_l+1}\M^{-k,j},~~~~~~~~~~~~~~n\ge k>l>j\ge\cases{0~~~if
{}~~N=2n+1\cr -1~~~if~~N=2n \cr}
\ee
\be
[\M^{l-1,k},\M^{1-l,l}]_{q^{-1}}=q^{\r_l-1}\M^{lk}~~~~~~~~~~
[\M^{-l,l-1},\M^{-k,1-l}]_{q^{-1}}=q^{\r_l}\M^{-k,-l}~~~~~~~~~~~~~~~~~~~~~
2\le l<k\le n
\ee
\be
[\M^{0k},\M^{01}]=q^{-1}\M^{1k}~~~~~~~~~~[\M^{-10},\M^{-k0}]=\M^{-k,-1}
{}~~~~~~~~~~~~~~~~~~~~~~~1<k\le n~~~if~~N=2n+1.
\ee
\end{prop}
$Proof$. As an example we prove equation $(70)_1$.
First consider the case $n=k$. We note that
\be
[\M^{-jl},\M^{-lk}]_q=q^{-2}\Lambda^{-\f 12}_k\mu_{-k}
\left[[\M^{-j,l},D^{-l}]_q,(X\cdot X)_{k-1}\right]D^k-\mu_k^{-\f 12}X^k
[\M^{-j,l},D^{-l}]_q,
\ee
as $[(X\cdot X)_{k-1},\M^{-j,l}]=0$. But
\be
[\M^{-j,l},D^{-l}]_q=D^{-j}q^{\r_l}
\ee
as a consequence of the preceding Lemma and Proposition 1, therefore the
RHS of equation $(70)_1$ gives $q^{\r_l}\M^{-jk}$. Applying Proposition 1
$(n-k)$ times we prove formula $(70)_1$ in the general case. The proofs of
the other equations are similar. $\diamondsuit$

\begin{prop}
When $q\in {\bf R}$
\be
(\k^i)^*=\k^i,~~~~~~~~~~~(\M^{1-k,k})^*=q^{-2}\M^{-k,k-1}~~~~~~k\ge 2,
{}~~~~~~~~~~~~~~~~~\cases{(\M^{01})^*=q^{-\f 32}\M^{-10}~~~~~if~~N=2n+1\cr
(\M^{12})^*=q^{-2}\M^{-2,-1}~~~~~if~~N=2n\cr}
\ee
\end{prop}
$Proof$. The thesis can be proved by writing these $\k,\M$ generators in terms
of the $B,L$ ones as shown in Appendix A and by using the conjugation
relations (60). $\diamondsuit$

The following three propositions give the basic commutation relations among
the Chevalley generators. More relations for the other roots
can be obtained from these ones using the relations of Proposition 5.
In the following two propositions we assume that
$k\ge\cases{1~~~if~~N=2n+1 \cr 2~~~if~~N=2n\cr}$

\begin{prop}
\be
[\k^i,\M^{\pm (1-k),\pm k}]_a=0~~~~~~a=\cases{q^{\pm 2}~~if~i=k\le n \cr
q^{\mp 2}~~~if~~i=k-1 \cr 1~~~otherwise}~~~~~~~~~~~~~~~~~~
[\k^i,\M^{\pm 1,\pm 2}]_a=0~~~~~~~a=\cases{q^{\pm 2}~~if~i=1,2 \cr
1~~~otherwise}
\ee
\end{prop}
$Proof$: a trivial consequence of formula (46) and of the definition
of $\M,k$'s. $\diamondsuit$

\begin{prop} (commutation relations between positive and negative simple roots)
\be
[\M^{1-m,m},\M^{-k,k-1}]_a=0~~~~~~~~~~~~~~~a=\cases{q^{-1}~~~~m-1=k \cr
1~~~~if~~m-1>k\cr}
\ee
\be
[\M^{-m,m-1},\M^{1-k,k}]_a=0~~~~~~~~~~~~~~~
a=\cases{q~~~if~~m-1=k\cr 1~~~if~~m-1>k\cr}
\ee
\be
[\M^{12},\M^{-2,1}]=0~~~~~~~~~~[\M^{-1,2},\M^{-2,-1}]=0~~~~~~~~~~~
{}~~~~~~~~~~if~~~~~N=2n,
\ee
\be
\cases{[\M^{1-m,m},\M^{-m,m-1}]_{q^2}=q^{1+2\r_m}\f{1-\k^{m-1}(\k^m)^{-1}}
{q-q^{-1}}~~~~~~~~2\le m\le n\cr
[\M^{01},\M^{-1,0}]_q=q^{-\f
12}\f{1-(\k^1)^{-1}}{q-q^{-1}}~~~~~~~~if~~N=2n+1\cr
[\M^{12},\M^{-2,-1}]_{q^2}=q^{-1}\f{1-(\k^2\k^1)^{-1}}{q-q^{-1}}
{}~~~~~~~~if~~N=2n.\cr}
\ee
\end{prop}
$Proof$. Use equations (42),(46) and perform explicit computations.
$\diamondsuit$

\begin{prop}
(Serre relations)
\be
[\M^{1-m,m},\M^{1-k,k}]=0~~~~~~~~~[\M^{-m,m-1},\M^{-k,k-1}]=0~~~~~~~~~~~~~~~
m,k>0,~~~~ |m-k|>1
\ee
\be
[\M^{12},\M^{1-j,j}]=0~~~~~~~~~~~~~[\M^{-2,-1},\M^{-j,j-1}]=0
{}~~~~~~~~~~~~~~~~~j=2,4,5,...,n,~~~~~~~N=2n,
\ee
\be
[\M^{1+j-m,m-j} ,\M^{2-m,m}]_a=0=[\M^{-m,m-2},\M^{j-m,m-j-1}]_a
{}~~~~~~~~~~~~~~a=\cases{q~~~if~~j=0\cr q^{-1}~~~
if~~j=1\cr}~~~~~~~~m\ge 3
\ee
\be
\cases{[\M^{01},\M^{12}]_{q^{-1}}=0 \cr [\M^{-1,2},\M^{02}]_q=0\cr}
{}~~~~~~~~~
\cases{[\M^{-2,-1},\M^{-1,0}]_{q^{-1}}=0\cr [\M^{-2,0},\M^{-2,1}]_q=0\cr}
{}~~~~~~~~~~~~~~~~~~~~~~if~~N=2n+1,
\ee
\be
\cases{[\M^{12},\M^{13}]_{q^{-1}}=0\cr [\M^{-23},\M^{13}]_q=0,\cr}~~~~~~~~~
\cases{[\M^{-3,-1},\M^{-2,-1}]_{q^{-1}}=0\cr [\M^{-3,-1},\M^{-3,2}]_q=0,\cr}
{}~~~~~~~~~~~~~~~~~~~~~~if~~N=2n.
\ee
\end{prop}
$Proof$. Use the definitions (62), commutation and derivation relations for
the $X,D$ variables, equation (37) and perform explicit computations.
$\diamondsuit$

We collect below all the basic commutation relations characterizing
$U_q^3,U_q^4$.
Their algebras read respectively
\be
\cases{[(\k^1)^{\f 12},L^{01}]_q=0\cr
       [(\k^1)^{\f 12},L^{-10}]_{q^{-1}}=0\cr
       [\M^{01},\M^{-10}]_q=q^{-\f 12}\f{1-(\k^1)^{-1}}{q-q^{-1}}\cr}
\ee
and
\be
\cases{[(\k^1\k^2)^{\f 12},\M^{12}]_{q^2}=0\cr
       [(\k^1\k^2)^{\f 12},\M^{-2,-1}]_{q^{-2}}=0\cr
       [\M^{12},\M^{-2,-1}]_{q^2}=q^{-1}\f{1-\k^1\k^2}{q-q^{-1}}\cr}
{}~~~~~~~~~~~~~~~~~~~~~
\cases{[((\k^1)^{-1}\k^2)^{\f 12},\M^{-1,2}]_{q^2}=0\cr
       [((\k^1)^{-1}\k^2)^{\f 12},\M^{-2,1}]_{q^{-2}}=0\cr
       [\M^{-12},\M^{-2,1}]_{q^2}=q^{-1}\f{1-(\k^1)^{-1}\k^2}{q-q^{-1}}\cr}
\ee
\be
[L,L']=0~~~~~~~~~L=\M^{12},\M^{-2,-1},(\k^1\k^2)~~~~~~
L'=\M^{-12},\M^{-21}, (\k^1)^{-1}\k^2;
\ee
We see that $U_q^4$ is the direct sum of two (commuting) identical algebras,
(the ones in the $L$ and $L'$ generators respectively). This is no surprise,
since it preludes to the relation
$U_q^4\ap U_q(so(4))\ap U_q(su(2))\otimes U_q(su(2))$,
which we will prove in section 5.

\section{The Hopf algebra structure of $U_q^N$ and its identification}
{}~~~~

In this section we show that $U_q^N$ is an Hopf algebra, more precisely
that it is isomorphic to $U_{q^{-1}}(so(N))$.

A natural bialgebra structure can be associated to $U_q^N$ for the reason
that its elements satisfy some Leibnitz rule when
acting as differential operators on $Fun({\bf R}_q^N)$. A matched antipode
can be found in a straightforward way, so that $U_q^N$ acquires
a Hopf algebra structure.
As for the mentioned isomorphism, we will prove it by constructing an
invertible transformation from
the generators of $U_q^N$ to those of $U_{q^{-1}}(so(N))$, in
such a way that the commutation relations,coproduct, counit, antipode of
$U_q^N$ are mapped into the ones of $U_{q^{-1}}(so(N))$.

This means that the Hopf algebra $U_{q^{-1}}(so(N)$ admits
a representation on all of $Fun({\bf R}_q^N)$.

 The Hopf algebra $U_q(so(N))$ [1][2] is generated by
$X_i^+,X_i^-,H_i$ ($i=1,...,n$) satisfying the commutation relations
\be
\cases{
[H_i,H_j]=0,~~~~~~~~~~~~~~[H_i,X_j^{\pm}]=\pm (\alpha_i,\alpha_j) X_i^{\pm},\cr
[X_i^+,X_j^-]=\delta_{ij}\f{q^{H_i}-q^{-H_i}}{q- q^{-1}} \cr
\sum\limits_{t=1}^{m_{ij}}(-1)^t\left[\stackrel{m_{ij}}{t}\right]_{q^i}
(X_i^{\pm})^t X_j^{\pm}(X_i^{\pm})^{m_{ij}-t}=0~~~~~~~~~i\neq j,\cr}
\ee
where
\be
q^i=q^{(\alpha_i,\alpha_i)},~~~~~~~~~m_{ij}=1-\f{(\alpha_i,\alpha_j)}{(\alpha_i
,\alpha_i)}~~~~~~~~~~~~\left[\stackrel{m}{t}\right]_q:=\f{[m]_q}{[t]_q[m-t]_q}
\ee
and the ($n$ x $n$) matrix of scalar products between the simple roots
$\alpha_i$ is given by
\be
\Vert b_{ij}\Vert :=\Vert (\alpha_i,\alpha_j)\Vert=\left\Vert\begin{array}
{ccccccccc}
   1  & -1 &    &    & & &    &    &    \\
   -1 &  2 & -1 &    & & &    &    &    \\
      & -1 &  2 & -1 & & &    &    &    \\
      &    &  . &  . &.& &    &    &    \\
      &    &    &    & & &    &    &    \\
      &    &    &    & &.&  . &  . &    \\
      &    &    &    & & & -1 &  2 & -1 \\
      &    &    &    & & &    & -1 &  2
\end{array}\right\Vert
\ee
if $N=2n+1$ and by
\be
\Vert b_{ij}\Vert :=\Vert (\alpha_i,\alpha_j)\Vert=\left\Vert\begin{array}
{cccccccccc}
   2 &    & -1 &    &    & & &    &    &    \\
     &  2 & -1 &    &    & & &    &    &    \\
  -1 & -1 &  2 & -1 &    & & &    &    &    \\
     &    & -1 &  2 & -1 & & &    &    &    \\
     &    &    &  . &  . &.& &    &    &    \\
     &    &    &    &    & & &    &    &    \\
     &    &    &    &    & &.&  . &  . &    \\
     &    &    &    &    & & & -1 &  2 & -1 \\
     &    &    &    &    & & &    & -1 &  2
\end{array}\right\Vert
\ee
if $N=2n$. Moreover, when $q\in {\bf R}$ they also satisfy the
adjointness relations
\be
H_i^{\dag}=H_i~~~~~~~~~~~(X_i^+)^{\dag}=X_i^-.
\ee

The coproduct, counit, antipode $\Phi_q,\epsilon,\sigma_q$ are defined
respectively by
\be
\Phi_q(H_i)={\bf 1}\otimes H_i + H_i\otimes {\bf 1}~~~~~~~~~~~~
\Phi_q(X^{\pm}_i)=X^{\pm}_i\otimes q^{-\f{H_i}2} + q^{\f{H_i}2}\otimes
X^{\pm}_i
\ee
\be
\epsilon(X^{\pm}_i)=0,~~~~~~~~~~~~~~\epsilon(H_i)=0
\ee
\be
\sigma_q(H_i)=-H_i~~~~~~~~~~\sigma_q(X^{\pm}_i)=-q^{-\f{(\alpha_i,\alpha_i)}2}
X^{\pm}_i
\ee
on the generators and extended as algebra homomorphisms/antihomomorphisms.

Now we show that there exist closed commutation relations between the
generators
of $U_q^N$ and the coordinates $x^i$.
\begin{prop}
\be
[\k^i,x^h]_a=0,~~~~~~a=\cases{q^2~~~if~h=i>0\cr
q^{-2}~~~if~~h=-i<0\cr 1~~~otherwise\cr}
{}~~~~~~in~Dif\!f({\bf R}_q^N)
\ee
\end{prop}
$Proof$. One has just to write $x^h$ as functions of
$\chi^j,{\cal D}_j$ and use relations $(46)$. $\diamondsuit$

As for the commutation relations between roots $\M$'s and $x^i$, we write down
only the ones involving simple roots and their opposite (the other ones can
be obtained in the same way or using Proposition 5).
\begin{prop}
Let $m\ge\cases{1~~if~~N=2n+1 \cr 2~~if~~N=2n\cr}$. Then
\be
[\M^{1-m,m},x^i]_a=0~~~~~~~~~~~a=\cases{1~~~~if~~~|i|<m-1~~or~~
|i|>m \cr q^{-1}~~~~~~~~if~~i=1-m,m},
\ee
\be
[\M^{-m,m-1},x^i]_a=0~~~~~~~~~~~a=\cases{1~~~~if~~~|i|<m-1~~or~~
|i|>m \cr q~~~~~~~~~if~~i=-m,m-1},
\ee
\be
[\M^{1-m,m},x^{m-1}]_q=-q^{\r_m}x^m~~~~~~~~~~~~~~~
[\M^{1-m,m},x^{-m}]_q=q^{\r_m}x^{1-m},
\ee
\be
[\M^{-m,m-1},x^m]_{q^{-1}}=-q^{\r_m}x^{m-1}~~~~~~~~~~
[\M^{-m,m-1},x^{1-m}]_{q^{-1}}=q^{\r_m}x^{-m},
\ee
\be
[\M^{01},x^0]=-q^{-1}x^1~~~~~~~~~~[\M^{-1,0},x^0]=x^{-1}~~~~~~~~~~~~~~~~~
if~~~N=2n+1
\ee
\be
\cases{[\M^{12},x^1]_{q^{-1}}=0~~~~~~~~~~~~~~~~~~~~[\M^{-2-1},x^{-1}]_q=0 \cr
       [\M^{12},x^2]_{q^{-1}}=0~~~~~~~~~~~~~~~~~~~~[\M^{-2-1},x^{-2}]_q=0 \cr
[\M^{12},x^{-1}]_q=-q^{-1}x^2~~~~~~~~~~~~~~[\M^{-2-1},x^1]_{q^{-1}}=q^{-1}x^{-2}
\cr [\M^{12},x^{-2}]_q=q^{-1}x^1~~~~~~~~~~~~~[\M^{-2-1},x^2]_{q^{-1}}=-q^{-1}
x^{-1}\cr}~~~~~~~~~~~~~~~~~~~~~if~~~N=2n.
\ee
\end{prop}
$Proof$. One has just to write $x^h$ as functions of
$\chi^j,{\cal D}_j$ and use relations (42),(46). $\diamondsuit$

The commutation relations of $\k^i,\M^{ij}$'s with $\p_i,\bar \p_i$ are the
same, since $\p_i\propto [\p\cdot\p,x_i]_{q^2}$,
$\bar\p_i\propto [\bar\p\cdot\bar\p,x_i]_{q^{-2}}$ and the $\k,\M$'s commute
with
scalars. The knowledge of the latter
commutation relations will allow us to construct
the inhomogeneous extension of $U_q^N\ap U_{q^{-1}}(so(N))$, i.e. the universal
enveloping algebra of the quantum Euclidean group, adding derivatives as new
generators [6].

We can consider $\M^{ij}$'s, $\k^i$'s as differential operators on
$Fun({\bf R}_q^N)$ in the same way as we did in section 2 with $\p_i$.
The commutation relations (97)-(103) allow us to define iteratively their
evaluations and Leibnitz rules starting from
\be
\k^i {\bf 1}|=1~~~~~~~~~~~~~~\M^{ij}{\bf 1}|=0
\ee
(${\bf 1}$ denotes the unit of $Fun({\bf R}_q^N$)). For instance, by applying
$\k^i$ to $x^h$, using eq. (97) and the previous relation we find
\be
\k^ix^h|=x^h\cdot\cases{q^{\pm 2}~~~if~~\pm h=i>0\cr 1~~~otherwise;\cr}
\ee
by applying $\k^i$ to $x^h\cdot g$ ($g\in Fun({\bf R}_q^N)$)
and using again eq. () we find
\be
\k^i(f\cdot g)|=\k^if|\k^ig|
\ee
for $f=x^h$ first, and then by recurrence for any $f\in Fun({\bf R}_q^N)$.
The latter relation is the Leibnitz rule for $\k^i$. (104),(105),(106) are
equivalent to (97),(104) and
determine the evaluation of $\k^i$ on all of $Fun({\bf R}_q^N)$.
Similarly the Leibnitz rule for the simple roots is determined
to be
\be
\M^{1-m,m}(f\cdot g)|=\M^{1-m,m}f|g+
(k^{m-1}(k^m)^{-1})^{\f 12}f|\M^{1-m,m}g|~~~~~~~~~~~~
m\ge\cases{1~~~~if~~N=2n+1\cr 2~~~~if~~N=2n\cr}
\ee
\be
\M^{12}(f\cdot g)|=\M^{12}f|g+(\k^1\k^2)^{-\f
12}f|\M^{12}g|~~~~~~~~~~~~if~~N=2n
\ee
($f,g\in Fun({\bf R}_q^N$)), and the same formulae hold by replacing each
simple
root by its negative partner.

More abstractly, the above formulae define: 1) a counit
$\epsilon: U_q^N\rightarrow {\bf C}$, by setting
$\epsilon(u):=\pi(u {\bf 1}|),~~~~~~u\in U_q^N$  and
$\pi(\alpha{\bf 1}):=\alpha~~~~\forall \alpha\in {\bf C}$, implying
that $\epsilon$ is an homomorphism
which on the generators $\k^i,\M^{ij}$ takes the form
\be
\epsilon(\M^{ij})=0~~~~~~~~~~~~~~~~~~~~\epsilon(\k^i)=1;
\ee
2) a coassociative
coproduct $\phi:U_q^N\rightarrow U_q^N\bigotimes U_q^N$
which on the generators $\k^i,\M^{ij}$ takes the form
\be
\phi(\k^i)=\k^i\bigotimes \k^i
\ee
\be
\cases{\phi(\M^{1-m,m})=\M^{1-m,m}\bigotimes
{\bf 1'}+ (\k^{m-1}(k^m)^{-1})^{\f 12}\bigotimes \M^{1-m,m} \cr
\phi(\M^{-m,m-1})=\M^{-m,m-1}\bigotimes
{\bf 1'}+ (k^{m-1}(k^m)^{-1})^{\f 12}\bigotimes \M^{-m,m-1}
\cr}~~~~~~~~~~~~~~~~
m\ge\cases{1~~~~if~~N=2n+1\cr 2~~~~if~~N=2n\cr}
\ee
\be
\cases{\phi(\M^{12})=\M^{12}\bigotimes {\bf 1'}+(\k^1\k^2)^{-\f 12}\bigotimes
\M^{12} \cr
\phi(\M^{-2,-1})=\M^{-2,-1}\bigotimes {\bf 1'}+(\k^1\k^2)^{-\f 12}\bigotimes
\M^{-2,-1} \cr}~~~~~~~~~~~~~~~~~if~~N=2n
\ee
(${\bf 1'}$ here denotes the unit of $Dif\!f({\bf R}_q^N)$, which acts
as the identity when considered as an operator on $Fun({\bf R}_q^N$),
and $\k^0\equiv {\bf 1'})$,
and is extended to all of $U_q^N$ as an homomorphism.
$\epsilon, \phi$ are matched so as to form a bialgebra; in particular
the coassociativity of $\phi$ follows from the associativity of the Leibnitz
rule, which in turn is a consequence of the associativity of
$Dif\!f({\bf R}_q^N$).

An antipode $\sigma$ which is matched with $\phi,\epsilon$, (i.e.
satisfies all the required axioms) is found by first imposing the two
basic axioms
\be
m\circ(\sigma\bigotimes id)\circ \phi=
m\circ(id\bigotimes \sigma)\circ \phi=i\circ \epsilon
\ee
on the generators of $U_q^N$, and then by extending it as an antihomomorphism;
here $m$ denotes the multiplication in $U_q^N$ and $i$ is the canonical
injection $i:{\bf C}\rightarrow U_q^N$.
Computations are straightforward:
\be
\sigma(\k^i)=(\k^i)^{-1},
\ee
\be
\cases{\sigma(\M^{1-m,m})=-(\k^m(\k^{m-1})^{-1})
^{\f 12}\M^{1-m,m}\cr
\sigma(\M^{-m,m-1})=-(\k^m(\k^{m-1})^{-1})
^{\f 12}\M^{-m,m-1}\cr}
{}~~~~~~~~~~~m\ge\cases{1~~~~if~~N=2n+1\cr 2~~~~if~~N=2n\cr}
\ee
\be
\cases{\sigma(\M^{12})=-(\k^1\k^2)^{-\f 12}\M^{12} \cr
\sigma(\M^{-2,-1})=-(\k^1\k^2)^{-\f 12}\M^{-2,-1}\cr}~~~~~~~~~~~if~~N=2n.
\ee

Finally, when $q\in {\bf R}$ it is straightforward to check that the complex
conjugation $*$ (the antilinear involutive antihomomorphism defined
in section 3, which acts on the
basic generators as shown in formula (75)) is compatible with the Hopf algebra
structure of $U_q^N$, so that $U_q^N$ gets a $*$-Hopf algebra.

Now it is easy to identify the Hopf algebra $U_q^N$.

\begin{prop}
All the relations characterizing the
($*$)-Hopf algebra $U_{q^{-1}}(so(N))$ are mapped into the ones characterizing
the ($*$)-Hopf algebra $U_q^N$ through the transformation of generators
$$
[\k^i(\k^{i-1})^{-1}]^{\f 12}=q^{H_i}~~~~~~~~~~~
\M^{1-i,i}= q^{\r_i-\f 32} X_i^+q^{-\f{H_i}2}~~~~~~~~~~~
\M^{-i,i-1}=q^{\r_i+\f 32}X_i^-q^{-\f{H_i}2},
$$
($i\ge\cases{1~~~if~~N=2n+1 \cr 2~~~if~~N=2n \cr}$, $\k^0\equiv 1$), and
\be
[\k^2(\k^1)]^{\f 12}=q^{H_1}~~~~~~~
\M^{1,2}= q^{-\f 52} X_1^+q^{-\f{H_1}2}~~~~~~~~
\M^{-2,-1}=q^{\f 12}X_1^-q^{-\f{H_1}2},~~~~~~~~~~~~~N=2n.
\ee
after setting $\Phi_{q^{-1}}=\phi$, $\sigma_{q^{-1}}=\sigma$
(or, alternatively, $\tau\circ\Phi_q=\phi$, $\sigma_q=\sigma$,
$\tau$ being the permutation operator).
In other words $U_q^N\ap U_{q^{-1}}(so(N))$.
\end{prop}
$Proof$. Straightforward computations. $\diamondsuit$

Note that if we had defined the elements of $U_q^N$ as differential operators
acting on $Fun({\bf R}_q^N)$ from the right (instead of from the left), we
would
have got the isomorphism $U_q^N\ap U_q(so(N))$.

As a concluding remark,
the final lesson we learn is that the product in $Fun({\bf R}_q^N)$ realizes
the tensor product
of representations of $U_{q^{-1}}(so(N))$, the Leibnitz rule satisfied
by the differential operators of $U_q^N$ realizes the corresponding coproduct,
and the real structure of $Dif\!f({\bf R}_q^N)$ realizes the real structure
of $U_{q^{-1}}(so(N))$.

\section{Representations}
{}~~~~

{}~~~Let us now look at $U_q^N$ as an operator algebra over $Fun({\bf R}_q^N)$.
In other words we consider `` evaluations '' of its elements on
$Fun({\bf R}_q^N)$ as defined in the previous section.
We look for its irreducible representations. Since $L\cdot L$ commutes with
any $L^{ij}$, it is proportional to the identity matrix on the base space
$W$ of each of them.

{}~~~~ As a first remark, we note that any $W$ must consist of polynomials
of fixed degree in $x$, as any $u\in U_q^N$ is a power series in
the products $x^i\p_j$. Of course, the degree of these polynomials must
be the same, say $k$, also after factoring out all powers of $x\cdot x$, since
$[u,x\cdot x]=0$.
One can easily realize (see Ref. [12]) that
the subspace of $Fun({\bf R}_q^N)$ satisfying these two requirements is
\be
W_k:=Span_{\bf C}[{\cal P}_{k,S~i_1...i_k}^{~~~~j_1...j_k}x^{i_1}...x^{i_k}],
{}~~~~~~~~~k\in {\bf N},
\ee
and that $W_k$ is an eigenspace of $L\cdot L$.
Here ${\cal P}_{k,S}$ denotes the (q-deformed)
$k$-symmetric (modulo trace) projector, which can
be defined through
\be
{\cal P}_{k,S}{\cal P}_{A~i,(i+1)}=0={\cal P}_{k,S}{\cal P}_{1~i,(i+1)},
{}~~~~~~~~~1 \le i \le k-1,
\ee
where ${\cal P}_{A~i,(i+1)}=(\otimes {\bf 1})^{i-1}\otimes{\cal P}_A\otimes
(\otimes {\bf 1})^{n-i-1}$, etc. Hence $W\subset W_k$.

In particular the fundamental (vector) representation $W_1$ is spanned by
the $N$ independent vectors $x^i$.

Below we are going to see that the representations of $U_q^N$ in $W_k$'s are
irreducible and of highest weight type. When $q=1$ they reduce to
the vector representations of $so(N)$.

As `` ladder operators '' corresponding to
positive, negative, simple roots
we take the ones indicated in Remark 1 for the case $q=1$. Correspondingly,

\begin{prop}
The highest (respectively lowest) weight eigenvector is
the vector $u_k^n:=(x^n)^k$ (respectively $(x^{-n})^k$). $W_k$ is generated by
iterated application of negative (resp. positive) ladder operators and is
an eigenspace of $L\cdot L$ with eigenvalue
\be
l^2_{k,N}=[k]_q[k+N-2]_q \f{(q^{\r_n+1}+q^{-\r_n-1})}
{(q+q^{-1})(q^{\r_n}+q^{-\r_n})}
\ee
\end{prop}
$Proof$. Using the derivation rules (42) it is straightforward to show
that all positive ladder operators $\M^{jk}$ annihilate $(x^n)^k$.
Moreover, it is easy to show that this vector is an eigenvector of $L\cdot L$
(with eigenvalue $l^2_k$) and therefore belongs to $W_k$.
This follows from the fact that it is an eigenvector of $x^l\p_l$
(with eigenvalue $(k)_{q^2}$) and from formulae (56),(30).
As already noted, the application of negative ladder operators then
yields a space $W \subset W_k$. As known, $W=W_k$ when $q=1$; but $dim(W),
dim(W_k)$ are constant with $q$, therefore $W=W_k$ $\forall q$. Similarly
one proves that $(x^{-n})^k$ is the lowest weight eigenvector.
$\diamondsuit$.

Let us consider the space of homogeneous polynomials of degree $k$
\be
M_k:=Span_{\bf C}[x^{i_1}...x^{i_k}].
\ee
As a consequence of the definition of $W_l$, we are able to
decompose $M_k$ into irreducible representations of $U_q^N$ (see [12]),
just as in the case $q=1$:
\be
M_k=\bigoplus\limits_{0\le m\le \f{k}{2}} W_{k-2m} (x\cdot x)^m.
\ee
Recall that $dim(M_k)={N+k-1 \choose N-1}$, therefore this formula allows
to recursively find $dim(W_k)$: $dim(W_k)=dim(M_k)-dim(M_{k-2})$.
The formula
\be
Fun({\bf R}_q^N)=\bigoplus\limits_{l=0}^{\infty}M_l=
\bigoplus\limits_{l=0}^{\infty}\bigoplus\limits_{0\le m\le \f l2}W_{l-2m}
(x\cdot x)^m,
\ee
gives the formal decomposition of $Fun({\bf R}_q^N)$ into irreducible
vector representations of $U_{q^{-1}}(so(N))$. All of
them are involved (infinitely many times),
and therefore $Fun({\bf R}_q^N)$ can be called the base
space of the `` regular '' representation $U_{q^{-1}}(so(N))$,
in analogy with the classical case.

When $q\in {\bf R}$,
starting from the prescriptions $(u_k^n,u_k^n):=1$,
$u^{\dag}:=u^*~~~~\forall u\in U_q^N$,
and using the commutation relations of $U_q^N$ one can define an inner
product $(~\cdot~,~\cdot~)$ in all of $W_k$.

\begin{prop}
The inner product $(~\cdot~,~\cdot~)$ is positive definite,
i.e. the representations $W_k$ are unitary (when $q\in {\bf R}^+$) w.r.t.
it.
\end{prop}
$Proof$. In Ref. [12] (or [14]) the integration $\int$ over
 ${\bf R}^N_q$ satisfying
Stoke's theorem was defined. According to it
\be
(f,g)=c_k\int dV~f^*g ~\r(x\cdot x),~~~~~~~~~f,g\in W_k.
\ee
Indeed, integrating by parts `` border terms '' vanish, and therefore taking
the adjoint $u^{\dag}$ of $u$ w.r.t this inner product amounts to taking its
complex conjugate.
In this formula $\r(x\cdot x)$ denotes a `` rapidly decreasing function ''
function of the square lenght such as the q-deformed gaussian
$exp_q(-ax\cdot x)$ and the normalization factor $c_k$ is chosen
so that $(u_k^n,u_k^n)=1$. But we have proved in Ref. [12] Lemma 7.3
that $(~\cdot~,~\cdot~)$ is positive definite. $\diamondsuit$

According to the theory of representations of $U_{q^{-1}}(so(N))$, when $q\in
{\bf R}$
the Cartan subalgebra generators $H_i$ make up
a complete set of commuting observables in $W_k$, $\forall k\ge 0$.
The highest weight associated to $W_k$ is the $n$-ple $(0,0,...,0,k)$
of eigenvalues of the $n$-ple of operators ($H_1,H_2,...,H_n)$ on $u_k^n$.

According to the commutation relations (76), a basis $E_k$ of $W_k$ consisting
of eigenvectors of $(H_1,H_2,...,H_n)$ is obtained by considering all the
independent vectors obtained by applying negative root operators to $u_k^n$.

For instance, in the case $N=3$ the dimension of $W_k$ is $2k+1$ and
\be
E_k:=\{u_{k,h}:=(\M^{-10} )^{-k-h}u_k^1,~~~h=-k,-k+1,...,k\}.
\ee
For any monomial
$M(k,\{i\}):=x^{i_1}x^{i_2}...x^{i_k}$ define $t(M):=i_1+i_2+...i_k$.
Looking at formulae (42) we realize that
the effect of the action of $\M^{0\pm 1}$ on any monomial $M_{\{i\}}$ is to
give a combination of monomials $M'$ with $t(M')=t(M)\pm 1$. Therefore
$u_{k,h}$ is a combination of monomials $M$ with $t(M)=h$.

The functions $(x\cdot x)^\f{-k}{ 2}u$  ($u\in E_k$)
will be said q-deformed spherical functions of degree $k$,
since they reduce to the classical ones in the limit $q=1$,
when we express $x^i(x\cdot x)^{\f{-1}{ 2}}$
in terms of angular coordinates.

\section{Appendix A}
{}~~~~

In this appendix we show how to express the generators $k_i,\M^{ij}$
as functions of $L^{ij},B_n$. One can easily check that this map is
invertible.

We first introduce some useful combinations $F$
of the $L^i,B$ variables introduced in subsection 3.1.

Let us iteratively define objects
${\cal F}_n^l\in Dif\!f({\bf R}_q^N)$, ($N=\cases{2n+1~~~for~~odd~N\cr 2n~~~
for~even~N\cr}$ as usual) by
\be
{\cal F}^{l+1}_l:={\cal B}_l,~~~~~~~l\ge 0~~~\forall N\ge 2;~~~~~~~~~~~
{\cal F}^{-1}_1=\mu_{-1}~~~~if~~N=2
\ee
(${\cal B}_0=1$ when $N=2$),
\be
{\cal F}_n^{l+1}(x,\p):=\mu_n{\cal F}_{n-1}^{l+1}(X,D),~~~~~
n>l\ge 0;~~~~~~~~~~~~~~
{\cal F}_n^{-1}(x,\p):=\mu_n{\cal F}_{n-1}^{-1}(X,D)~~~~~~~~~if~N=2n.
\ee
Let $F_n^l:={\cal F}_n^l\Lambda^{-\f 12}$. One easily checks that
$F_n^l\in U_q^N$, more precisely
\be
F_n^{l+1}=B_n+\f{q^2-1}{1+q^{-2\r_n}}
[\sum\limits_{j=l+1}^nL^jq^{-\r_j} - (q^2-1)\f{(n-l)_{q^2}}{1+q^{2\r_l}}
\sum\limits_{j=1}^l L^jq^{-\r_j}]
\ee
\be
F^{-1}_n=B_n+\f{q^2-1}{1+q^{-2\r_n}}
\sum\limits_{j=1}^nL^jq^{-\r_j}+(1-q^2)L^1=F^1_n
+(1-q^2)L^1=F^2_n - \f{q^2-1}{2}L^1.
\ee
Next we define
\be
K^{i+1}_n:=(\Lambda_n)^{-1}\Lambda_i(\mu_{i+1})^2...(\mu_n)^2\in
Dif\!f({\bf R}_q^N),~~~~~~~~0\le i\le n-1
\ee
and observe that
\begin{prop}
$K_n^i$'s belong to $U_q^N$ and
\be
K^{i+1}_n=(F^{i+1}_n)^2 -\f{q^2-1}{q^{2\r_i}+1}\f{q^2-q^{-2}}{1+q^{-2\r_i-2}}
(L\cdot L)_i;
\ee
\end{prop}
$Proof.$ From the definitions of $\Lambda_l,\mu_l$ it immediately follows
the first part of the proposition. Relation (131) is a consequence of formula
(56) and of the definitions of the $F$'s.$\diamondsuit$

Formulae (128),(129),(131) allow to express $\k_i$ as functions
of $L^{ij},B$ after noting that
\be
\k_i=K^i_n(K_n^{i+1})^{-1}~~~~~~~~~~~~~~~~~~~~~(K^{n+1}_n\equiv 1)
\ee
As for the $\M$'s, we find the
\begin{prop}
\be
\cases{\M^{1-k,k}=(K^k_n)^{-1}\left[F^{k-1}_nL^{1-k,k}-\f{q^2-1}{q^2+
q^{-2-2\r_k}}\sum\limits_{l=2-k}^{l=k-2}L^{1-k,l}L_l^{~~k}\right] \cr
\M^{-k,k-1}=q^{-1}(K^k_n)^{-1}\left[L^{-k,k-1}F^{k-1}_n-\f{q^2-1}{q^2+
q^{-2-2\r_k}}\sum\limits_{l=2-k}^{l=k-2}L^{-k,l}L_l^{~~k-1}\right]}
{}~~~~~~~~~~~~~~~2\le k\le n
\ee
and
\be
\cases{\M^{0\pm 1}=(F^1_n)^{-1}L^{0\pm 1}~~~~~~~~~if~~N=2n+1\cr
\M^{\pm(1,2)}=(F^1_n)^{-1}L^{\pm(1,2)}~~~~~~~~~~~~if~~N=2n\cr}.
\ee
\end{prop}
$Proof$. As an example we prove equation $(133)_1$. As usual, it is sufficient
to prove the claim when $k=n$, and then use Proposition 1 to extend it to
$n>k$. Inverting relation $(33)_4$ we get
$D^n=q\Lambda^{-\f 12}\mu_n^{\f 12}[\p^n+q^{-2-2\r_n}(q^2-1)X^n(D\cdot D)
_{n-1}]$.
Replacing this expression in the definition (62) of $\M^{1-n,n}$ and
using the definition (27) for $\Lambda_{n-1}$ we easily find
$$
\Lambda_n\M^{1-n,n}=\mu_{-n}\mu_n^{\f 12}\left\{[D^{1-n},(X\cdot X)_{n-1}]
\p^n-[D^{1-n},(1+q^{-2-2\r_n})(X\cdot D)_{n-1}]X^n\right\}
$$
$$
=\mu_{-n}\mu_n^{\f 12}\{\mu_{n-1}(q^{2\r_n+2}X^{1-n}\p^n-X^nD^{1-n})+
(1-q^{-2})[(X\cdot X)_{n-2}D^{1-n}\p^n
$$
\be
+X^{1-n}(D\cdot D)_{n-2}X^n]-(q^2-1)(1+q^{-2\r_n-4})(X\cdot D)_{n-2}]
X^nD^{1-n}\};
\ee
on the other hand, using the normalization (54) for ${\cal L}^{ij}$,
$$
\f{\sum\limits_{l=2-n}^{n-2}{\cal L}^{1-n,l}{\cal L}_l^{~~n}}{1+q^{-2\r_n-4}}=
(\p\cdot x)_{n-2}x^{1-n}\p^n-(x\cdot x)_{n-2}\p^{1-n}\p^n-x^{1-n}(\p\cdot \p)
_{n-2}x^n+q^2(x\cdot \p)_{n-2}\p^{1-n}x^n
$$
$$
=\mu_n^{\f 32}\{(X\cdot D)_{n-2}(q^2D^{1-n}X^n+q^{-2-2\r_n}X^{1-n}\p^n)+
\f{q^2-q^{2\r_n+4}}{q^2-1}\mu_{n-1}X^{1-n}\p^n
$$
\be
-(X\cdot X)_{n-2}D^{1-n}\p^n-X^{1-n}(D\cdot D)_{n-2}X^n\}
\ee
and
\be
{\cal F}_n^{n-1}{\cal L}^{1-n,n}=\mu_n^{\f 32}\left[1+(q^2-1)(X^{n-1}D_{n-1}+
q^{-2\r_n-4}(X\cdot D)_{n-2})\right](X^{1-n}\p^n-X^nD^{1-n}).
\ee
{}From the preceding three formulae we find that
\be
\Lambda_n\M^{1-n,n}=\mu_{-n}(\mu_n)^{-1}[{\cal F}_n^{n-1}{\cal L}^{1-n,n}-
\f{q^2-1}{q^2+q^{-2-2\r_n}}\sum\limits_{l=2-n}^{l=n-2}{\cal L}^{1-n,l}
{\cal L}_l^{~~n}]
\ee
which is equivalent to the claim upon use of formula (132). $\diamondsuit$

Note that $K^1_n=(F^1_n)^2$ both for odd and even $N$, and
$(K^2_n)^2=K^1_n(F^{-1})^2$ when $N=2n$. All $K^i_n$ go to 1 in the limit
$q\rightarrow 1$.
Moreover, for $N=3$ $F^1_1=(\k^1)^{\f 12}$ and for $N=4$
$F^1_2=(\k^1\k^2)^{\f 12}$
$F^{-1}_2=(\k^1)^{-\f 12}(\k^2)^{\f 12}$.

\section{Appendix B}

Define
\be
\cases{\hat\M^{in}:=X^iD^n-q^{-2-2\r_n}\mu_n^{\f 12}\Lambda_n^{-\f 12}
[X^i,(D\cdot D)_{n-1}]X^n \cr
\hat\M^{-n,i}:=X^{-n}D^i-q^{-3-2\r_n}\Lambda_N^{-\f 12}\mu_n^{-\f 12}\mu_{-n}
[D^i,(X\cdot X)_{n-1}]D^{-n} \cr}~~~~~~~~~~~|i|<n.
\ee
\begin{lemma}

$\hat \M^{in},\hat \M^{-n,i}\in U_q^N$ and can be easily expressed as simple
functions of the $\M,\k$'s.
\end{lemma}

Since $[\k^n,\chi^j]_a=0=[\k^n,\Dr_j]_b$ with some $a,b$, we can
introduce a grading $p\in {\bf Z}$ in $Dif\!f({\bf R}_q^N)$ and
decompose the latter as follows
\be
Dif\!f({\bf R}_q^N)=\bigoplus\limits_{p\in {\bf Z}}Dif\!f^p~~~~~~~~~~~~~~~
where~~~~~~\k^nDif\!f^p:=q^{2p}Dif\!f^p\k^n;
\ee
note that for each monomial
$M(\chi,\Dr):=(\chi^n)^l(\chi^{-n})^m(\Dr_n)^s(\Dr_{-n})^r$
\be
p(M)=l+r-m-s.
\ee
Decomposition (140) induces the decomposition
$U_q^N=\bigoplus\limits_{p\in {\bf Z}}U_q^N\bigcap Dif\!f^p$.

Now we can sketch the proof of the main theorem of this appendix.
\begin{prop}
\be
u\in U_q^N~~~~~~~~~~~~~~~\Rightarrow~~~~~~~~~~~~~
{}~~~~u=u(\k^i,\M^{jk}),~~~~~i=1,...,n,~~~~~~~~~~|j|,|k|\le n.
\ee
Moreover
\be
f\in Dif\!f({\bf R}_q^N):~~~~\left[f,\cases{x\cdot x \cr \p\cdot
\p\cr}\right]=0
{}~~~~~~~~~~~~~~\Rightarrow~~~~~~~~~~~~~~~
f=\sum_lu_l\cases{f_l(x)\cr f_l(\p)\cr},~~~~~u_l\in U_q^N.
\ee
\end{prop}
$Sketch~of~the~Proof$. As a preliminary remark, let us recall that
$[\Lambda_n,u]=0$,
namely $u$ has natural dimension zero. Our proof will be by induction in $n$.
It is
easy to prove that $U_q^1={\bf 1}\cdot{\bf C}$, and that $U_q^2$ is generated
by
$\k^1$. Now assume that the thesis is true for $U_q^{N-2}$.

The most general $u\in Dif\!f({\bf R}_q^N)$ can be written in the form
$$
u=\sum\limits_{l,m=0}^{\infty}\{(\chi^{-n})^l(\mu_n^{-\f 12}\chi^n)^m
v_{l,m}(\mu_n,\mu_{-n},\chi^j,\Dr_j)+(\mu_n^{-\f 12}\Dr_n)^l(\Dr_{-n})^m
v_{-l,-m}(\mu_n,\mu_{-n},\chi^j,\Dr_j)+
$$
\be
(\Dr_{-n})^l(\mu_n^{-\f 12}\chi^n)^m v_{-l,m}(\mu_n,\mu_{-n},\chi^j,\Dr_j)+
(\chi^{-n})^l(\mu_n^{-\f 12}\Dr_n)^m v_{l,-m}(\mu_n,\mu_{-n},\chi^j,\Dr_j)+
\ee
$|j|<n$. In fact the dependence on powers of $\chi^{\pm n}\Dr_{\pm n}$ can be
reabsorbed into the dependence on $\mu_{\pm n}$.

It is easy to realize that, if we impose the constraint that
the natural dimension $d(u)$ of $u$ is zero, formula (144) can be rewritten
in the form
$$
u=\sum\limits_{\{l_i,l'_i\}}
(\M^{-n,1-n})^{l_{1-n}}....(\M^{-n,n-1})^{l_{n-1}}(\M^{1-n,n})^{l'_{1-n}}...
(\M^{n-1,n})^{l'_{n-1}}\cdot
$$
\be
\sum\limits_{p=0}^{\infty}\left[
\sum\limits_{h=0}^p[(\mu_n^{-\f 12}\chi^n)^h(\Dr_{-n})^{p-h}
v^{p,h}_{\{l_i,l'_i\}}(\mu_n,\mu_{-n},\chi^j,\Dr_j)
+(\mu_n^{-\f 12}\Dr_n)^h(\chi^{-n})^{p-h}
v^{-p,-h}_{\{l_i,l'_i\}}(\mu_n,\mu_{-n},\chi^j,\Dr_j)]\right],
\ee
where $i=1-n,...,n-1$.
We sketch the procedure which leads to this result.
For each $\mu_n^{-\f 12}\chi^n$ or $\chi^{-n}$ (respectively $\Dr^{-n}$ or
$\mu_n^{-\f 12}\Dr^n$) we can extract out of the corresponding coefficient
function $v$ a $D^i$
(respectively a $X^i$) variable (since $d(u)=0$) and replace the LHS's of
the following identities by the RHS's (see the definitions (62)):
\be
\mu_n^{-\f 12}\chi^nD^i=q^{-2}\Lambda_{n-1}\k_n^{-\f 12}[D^i,
(X\cdot X)_{n-1}]\Dr^n-\M^{in},~~~~~~~~~
\chi^{-n}D^i=q^{-2}\Lambda_{n-1}\k_n^{-\f 12}[D^i,
(X\cdot X)_{n-1}]\Dr^n-\hat\M^{in}
\ee
\be
\mu_n^{-\f 12}\Dr^{-n}X^i=q^{-1}\Lambda_{n-1}\k_n^{-\f 12}
[(D\cdot D)_{n-1},X^i]\chi^{-n}-\M^{-n,i},~~~~~~~~
\chi^{-n}D^i=q^{-2}\Lambda_{n-1}\k_n^{-\f 12}[D^i,
(X\cdot X)_{n-1}]\Dr^n-\hat\M^{in}.
\ee
Then each factor $\chi^n\Dr_n$, $\chi^{-n}\Dr_{-n}$ can be reabsorbed into
the $\mu_n,\mu_{-n}$-dependence of the coefficient functions $v$'s. Finally,
we arrive at (145) using the result of Lemma 2 and the commutation relations
of section 3, which allow us to reorder all $\M,\k$'s according to the
ordering shown in that formula.

Now we impose the conditions $[u,x\cdot x]=0=[u,\p\cdot \p]$ explicitly.
They reduce to
\be
\cases{
\left[\sum\limits_{h=0}^p(\mu_n^{-\f 12}\chi^n)^h(\Dr_{-n})^{p-h}
v^{p,h}_{\{l_i,l'_i\}}(\mu_n,\mu_{-n},\chi^j,\Dr_j),\cases{x\cdot x
\cr \p\cdot\p\cr} \right]=0 \cr
\left[\sum\limits_{h=0}^p(\mu_n^{-\f 12}\Dr_n)^h(\chi^{-n})^{p-h}
v^{-p,-h}_{\{l_i,l'_i\}}(\mu_n,\mu_{-n},\chi^j,\Dr_j),\cases{x\cdot x
\cr \p\cdot\p\cr} \right]=0. \cr}
\ee
In fact the powers of $\M$'s appearing in formula (145) belong to a Poincare'
basis of $U_q^N$, therefore are independent,
and their coefficient functions can be split into components belonging
to different subspaces $Dif\!f^p$ (140).
Using a procedure which, for the sake of brevity, we describe only in the
case $p=1$,
it is easy to show that from the latter equations it follows decompositions
of the type
\be
\cases{
\sum\limits_{h=0}^p(\mu_n^{-\f 12}\chi^n)^h(\Dr_{-n})^{p-h}
v^{p,h}_{\{l_i,l'_i\}}(\mu_n,\mu_{-n},\chi^j,\Dr_j)=
\sum\limits_{\{i_1,...i_p\}}f_{\{i_1,...i_p\}}(\k^n)u_{\{i_1,...i_p\}}
\M^{i_1,n}...\M^{i_p,n},  \cr
\sum\limits_{h=0}^p(\mu_n^{-\f 12}\Dr_n)^h(\chi^{-n})^{p-h}
v^{-p,-h}_{\{l_i,l'_i\}}(\mu_n,\mu_{-n},\chi^j,\Dr_j)=
\sum\limits_{\{i_1,...i_p\}}f_{\{i_1,...i_p\}}(\k^n)u_{\{i_1,...i_p\}}
\M^{-n,i_1}...\M^{-n,i_p},  \cr}
\ee
$u_{\{i_1,...i_p\}}\in U_q^{N-2}$,
which completes the proof of formula (142). When $p=1$,
upon use of formulae (42)(46), it is easy to verify that the LHS's of
equations $(148)$ are combination of
$(\mu_n^{-\f 12}\chi^n)^2\chi^{-n},\Dr_{-n}$,
$\mu_n^{-\f 12}\chi^n$
and $\mu_n^{-\f 12}\Dr_n(\Dr_{-n})^2,\Dr_{-n},\chi^n$ respectively, and that
setting their coefficients equal to zero amounts to
\be
v^m=v^m(\k^n,\chi^j,\Dr_j)~~~~~~~~~m=0,1,~~~|j|<n~~~~~~~~~~~~~~~~~~~~~~~
[v^0,(X\cdot X)_{n-1}]=0=[v^1,(D\cdot D)_{n-1}]
\ee
\be
v^0=-q^{-2-2\r_n}\Lambda_{n-1}^{-\f 12}(\k^n)^{-\f 12}[v^1,(X\cdot X)_{n-1}].
\ee
Hence
\be
0=\left[[v^1,(X\cdot X)_{n-1}],(X\cdot X)_{n-1}\right]_{q^2}=
\left[[v^1,(X\cdot X)_{n-1}]_{q^2},(X\cdot X)_{n-1}\right]
\ee
implying upon use of the recursion hypothesis (143),formula (25) and of
relations $d(v^1)=1$,
\be
[(D\cdot D)_{n-1},(X\cdot X)_{n-1}]_{q^2}(q^2-1)=q^{4+2\r_n}(\Lambda_{n-1}-
q^{2\r_n}),
\ee
the equation
\be
[v^1,(X\cdot X)_{n-1}]_{q^2}=u_iX^i,~~~~~u_i\in U_q^{N-2}
{}~~~~~~~~~~~\Rightarrow~~~~~~~~~~~v^1 \propto \left[v^1,\f{q^{4+2\r_n}
(\Lambda_{n-1}-q^{2\r_n})}{q^2-1}\right]_{q^2}\propto u_iD^i.
\ee
This yields $v^1(\mu_n^{-\f 12}\chi^n)+v^0\Dr_{-n}\propto u_i\M^{in}$, as
claimed.

The proof of (143) can be given recursively
by constraining the general expansion (144) in a similar way. $\diamondsuit$.

\bc
{\large{\bf Acknowledgments}}
\ec

I thank L. Bonora, V. K. Dobrev, M. Schlieker and J. Wess for useful
discussions.

\section {\bf References}
{}~~~~

1. V. G. Drinfeld, " Quantum Groups ", Proceedings of the International
Congress of Mathematicians 1986, Vol. 1, 798; M. Jimbo, Lett. Math. Phys.
{\bf 10} (1986), 63.

{}~

2. L. D. Faddeev, N. Y. Reshetikhin and L. A. Takhtajan, " Quantization of Lie
Groups and Lie Algebras ", Algebra and Analysis, {\bf 1} (1989) 178, translated
from the Russian in Leningrad Math. J. {\bf 1} (1990), 193.

{}~

3. Yu. Manin, preprint Montreal University, CRM-1561 (1988); " Quantum
Groups and Non-commutative Geometry ", Proc. Int. Congr. Math., Berkeley
{\bf 1} (1986) 798; Commun. Math. Phys. {\bf 123} (1989) 163.

{}~

4. J. Wess and B. Zumino, Nucl. Phys. Proc. Suppl. {\bf 18B} (1991), 302;
W. Pusz and S. L. Woronowicz, Reports in Math. Phys. {\bf 27} (1990), 231.

{}~

5. G. Fiore, in preparation.

{}~

6. G. Fiore, forthcoming paper.

{}~

7. O. Ogievetsky, W. B. Schmidke, J. Wess and B. Zumino, Commun. Math.
Phys. {\bf 150} (1992) 495-518. See also: J. Wess, `` Differential calculus
on quantum planes and applications '', talk given on occasion of the third
centenary celebrations of the Mathematische Gesellschaft Hamburg, March 1990,
KA-THEP-1990-22.

{}~

8. M. Schlieker, W. Weich and R. Weixler, Z. Phys. {\bf C 53} (1992), 79-82;
S. Majid, J. Math. Phys. {\bf 34} (1993), 2045.
{}~

9. O. Ogievetsky, Lett. Math. Phys. {\bf 24} (1992),245.

{}~

10. U. Carow-Watamura, M. Schlieker and S. Watamura, Z. Phys. C Part. Fields
{\bf 49} (1991) 439.

{}~

11. O. Ogievetsky and B. Zumino, Lett. Math. Phys. {\bf 25} (1992),121.

{}~

12. G. Fiore, Int. J. Mod. Phys. {\bf A8} (1993), 4679.

{}~

13. G. Fiore, Nuovo Cimento {\bf 108 B}, 1427.

{}~

14. A. Hebecker and W. Weich, Lett. Math. Phys. {\bf 26} (1992), 245.

\end{document}